\documentclass[twocolumn,english,aps,amsmath,showpacs]{revtex4}
\usepackage{graphicx}
\usepackage{amsmath,amssymb,amsfonts}
\makeatletter

\providecommand{\LyX}{L\kern-.1667em\lower.25em\hbox{Y}\kern-.125emX\@}

\makeatother
\begin{document}

\title{ Spiking Optical Patterns and Synchronization}

\author {Michael Rosenbluh\( ^{1} \), Yaara Aviad\( ^{1} \),
Elad Cohen\( ^{1} \), Lev Khaykovich\( ^{1} \), Wolfgang Kinzel\(
^{2} \), Evi Kopelowitz\( ^{1} \), Pinhas Yoskovits\( ^{1} \) and
Ido Kanter\( ^{1} \) }

\affiliation{\( ^{1} \)Department of Physics, Bar-Ilan University,
Ramat-Gan, 52900 Israel,}

\affiliation{\( ^{2} \)Institut f\"ur Theoretische Physik,
Universit\"at W\"urzburg, Am Hubland 97074 W\"urzburg, Germany}

\begin{abstract}
We analyze the time resolved spike statistics of a solitary and two
mutually interacting chaotic semiconductor lasers whose chaos is
characterized by apparently random, short intensity spikes.
Repulsion between two successive spikes is observed, resulting in a
refractory period which is largest at laser threshold. For time
intervals between spikes greater than the refractory period, the
distribution of the intervals follows a Poisson distribution. The
spiking pattern is highly periodic over time windows corresponding
to the optical length of the external cavity, with a slow change of
the spiking pattern as time increases. When zero-lag synchronization
between the two lasers is established, the statistics of the nearly
perfectly matched spikes are not altered. The similarity of these
features to those  found in complex interacting neural networks,
suggests the use of laser systems as simpler physical models for
neural networks.

\end{abstract}


\maketitle

Semiconductor lasers, subjected to optical feedback, display chaotic
behavior \cite{1}. The chaotic behavior consists of a very short and
random spiking of the laser intensity with the time between spikes
depending on how far above lasing threshold the laser is. Two
chaotic lasers can be synchronized with each other and this has
allowed them to be excellent candidates for novel broadband
\cite{2,3,4,5,6} communication devices. Different configurations,
such as delayed optoelectronic \cite{7,8} or coherent optical
injection \cite{8,9,10,11} have been used for synchronization of the
two lasers. Using optical feedback, configurations consisting of
unidirectional \cite{7,8} or mutual coupling \cite{5,6,8,9,10,11}
and variations of the strength of the self and coupling feedback
have been shown to result in different synchronization states. The
lasers can synchronize in a leader-laggard or anticipated mode, as
well as in two different synchronization states; achronal or
generalized synchronization \cite{12,13,14} where the cross
correlation is time shifted by the feedback delay time but neither
laser acts as a preferred leader or laggard, or isochronal
synchronization (zero-lag) where there is no time delay between the
two lasers' chaotic signals \cite{5,6,15,16,17}.

Zero-lag synchronization of lasers was recently extended to a
cluster consisting of three semiconductor lasers, mutually coupled
along a line, in such a way that the central laser element acts as a
relay of the dynamics between the outer elements \cite{20,21}. The
zero-lag synchronized dynamics of remotely located chaotic signal
sources has sparked an interest in such systems in part because they
have features also seen in biological and neural transmission
networks.  Though the time scales for the two phenomena are vastly
different; lasers spiking on 100 ps time scale while neurons spike
on ms time scales, much of the dynamics and spiking statistics
appear to have common behavior. Here we report on the spiking
optical pattern of solitary and two mutually coupled chaotic lasers,
observed on a time scale which resolves the individual spikes,  in
both their synchronized and unsynchronized states and determine the
statistical behavior of the spiking. This further allows us to
establish an analogy to the spiking behavior of single and
interacting neurons.

Our experimental setup is shown schematically in Figure 1, where two
semiconductor lasers are coupled via a partially transmitting mirror
placed in the middle of the coupling optical path between the
lasers. In the actual experiment the self feedback and the mutual
feedback paths were spatially separated through the use of beam
splitters \cite{6}. The time for light propagation from the mirror
to one of the lasers is  $\tau/2$  . We distinguish between the
following three limiting scenarios. In the case of a fully
reflecting mirror, the lasers are uncoupled and each laser is
subject only to a delayed self-feedback. In the second scenario
where the mirror is fully transparent, each laser receives a delayed
signal from the other, and this configuration is known as
"face-to-face". In the intermediate scenario, the mirror is
partially transmitting and each laser is driven by two delayed
signals, one from self-coupling and one from mutual coupling.

\begin{figure}
\vspace{-0.5 cm} {\centering
\resizebox*{0.4\textwidth}{0.08\textheight}
{{\includegraphics[angle=0]{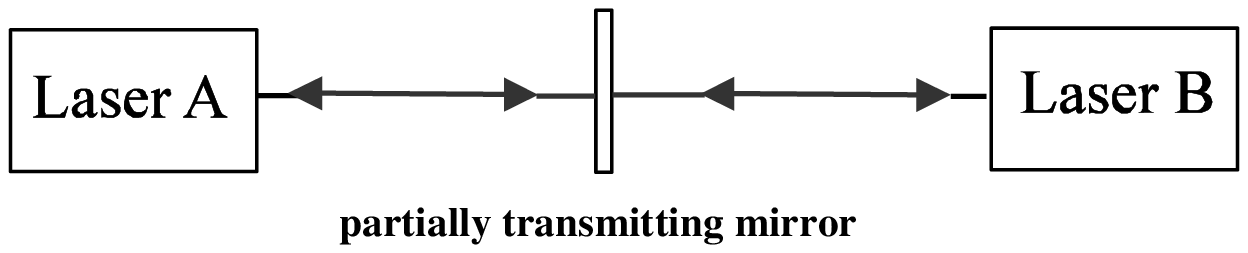}}}
\par}
\caption{Schematic experimental setup. Two lasers are mutually
coupled via a partially transmitting mirror placed in the middle of
the coupling optical path between the lasers. }
\end{figure}

In the experiment we used Fabry-Perot semiconductor lasers emitting
at $670$ nm wavelength, selected to have nearly the same threshold
current, emission wavelength, and output power. The temperature of
each laser is stabilized to better than $0.01K$ and the individual
laser temperatures are tuned so that the two lasers have nearly
identical output wavelengths. The self and mutual feedback loop time
is $23.55$ ns. Two fast ($50$ GHz bandwidth) detectors biased via a
$40$ GHz bandwidth bias T measure the output intensity of each
laser. The DC current into the bias T is used to measure the average
DC power falling on the detector while the AC currents are measured
simultaneously in two channels of a $12$ GHz bandwidth, $40$ GbGS/s
oscilloscope (Tektronix TDS 6124C).

For the case in which the mirror is fully reflecting the lasers are
decoupled. Each laser becomes chaotic due to the self-feedback but
their chaotic fluctuations are completely independent of each other.
A typical trace of one of the laser output intensity measurements is
presented in the top panel of Figure 2a, where the ratio of the
actual laser current to the threshold current, $p$, is $1.03$. The
time dependent intensity of the laser consists of spikes of typical
duration of $\sim120$ ps. In the following panels we show the same
laser intensity fluctuations recorded after a time, $\tau(=23.55$
ns) , $2\tau$ and $3\tau$. Figure 2 clearly indicates that on a time
scale of a few optical delayed self-feedback times the timing of the
spikes repeats itself, where as time elapses the spikes gradually
broaden and finally disappear as new spikes emerge in new positions,
forming a new pattern. This behavior is physically easy to
understand since the feedback photons circulate in the long external
cavity with a periodicity $\tau$ and the lasers receives nearly the
same feedback pattern with this periodicity.  The photon lifetime in
the cavity is finite, however, and thus the feedback waveform slowly
changes.  After many $\tau$ periods (many photon round trips) the
feedback waveform and the laser's chaotic fluctuation pattern will
have changed completely.  From this explanation we can also see that
the revival of the chaotic pattern with period $\tau$ will be nearly
independent of p since the chaos is caused and determined by the
feedback photon train which has a similar form with period $\tau$.
To further demonstrate the repeatability of the patterns after a
delay  $\tau$  in Figure 2b we show the overlaid intensity trace at
time $t+\tau$  (green) and at time $t+2\tau$  (blue).

\begin{figure}
\vspace{-0.5 cm} {\centering
\resizebox*{0.45\textwidth}{0.3\textheight}
{{\includegraphics[angle=0]{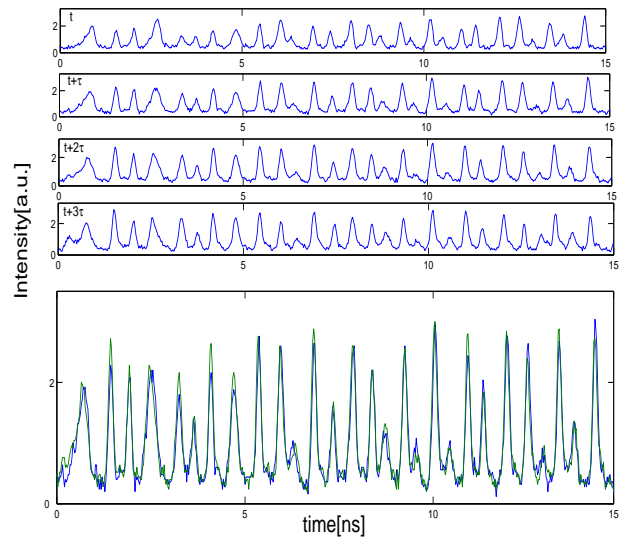}}}
\par}\caption{(a) A trace of $15$ ns duration of the intensity of one laser followed by
plots of the same laser intensity after a time   $\tau$, $2\tau$ and
$3\tau$  with $\tau =23.55$ ns. The laser was operating with
$p=1.03$ and with a reflected power of a few $\%$ of the laser
output intensity. (b) The intensity trace at time $t+\tau$ (green)
and at time $t+2\tau$  (blue) demonstrating the slowly decaying
periodicity of the spiking pattern. }
\end{figure}

A quantitative analysis of the spike statistics requires the
definition of a low-signal threshold so as to eliminate the small
spikes in the measured laser intensity which are the result of noise
in the measurements. Our threshold was chosen to be at twice the
average detector noise level, and the timing of a spike above this
threshold level was determined according to the time of its maximum
intensity. For each measurement we accumulated $70,000$ consecutive
spikes. A histogram of the time interval between consecutive spikes
is presented in Figure 3a for various values of $p$.

The probability distribution of the time intervals between two
spikes consists of two main features. The distribution for
relatively long time intervals between the spikes follows a random,
Poisson distribution where for small $p$ values the exponential
decay rate increases linearly from zero as a function of $p$, as
shown in Figure 3b4a. For very short time intervals the Poisson
distribution is altered so that immediately after a spike it is most
improbable to record a second spike. The most probable time between
consecutive spikes is defined, as for neural spike trains, as the
refractory time \cite{26,27}. Figure 3c 4b indicates that the
refractory time increases as the laser current approaches the
threshold value, which makes physical sense, since at low pumping
currents it takes longer to rebuild the laser gain after the
previous pulse had depleted it.

The chaotic dynamics are well described by the three coupled
Lang-Kobayashi rate equations for the optical field amplitude $E$,
the optical phase  $\Phi$   and the excited carriers, $n$, of the
gain medium \cite{28}. The equations also predict that the intensity
chaos is due to a spiking behavior of the lasers on a 100 ps time
scale, with no laser emission between spikes, in full agreement with
the observations.  Simulations From the simulations of the
Lang-Kobayashi equations indicate we also determined that after a
spike the laser ceases to emit because the number of carriers drops
well below threshold, hence a successive spike is forbidden until
the population is repumped again by the laser injection current.

\begin{figure}
\vspace{-0.5 cm} {\centering
\resizebox*{0.5\textwidth}{0.3\textheight}
{{\includegraphics[angle=0]{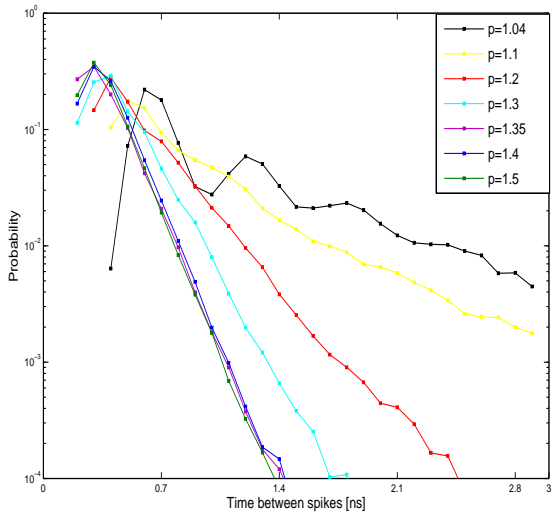}}}
\par}\caption{The probability for time intervals between spikes for various values
of $p$.}
\end{figure}

\begin{figure}
\vspace{-0.5 cm} {\centering
\resizebox*{0.5\textwidth}{0.3\textheight}
{{\includegraphics[angle=0]{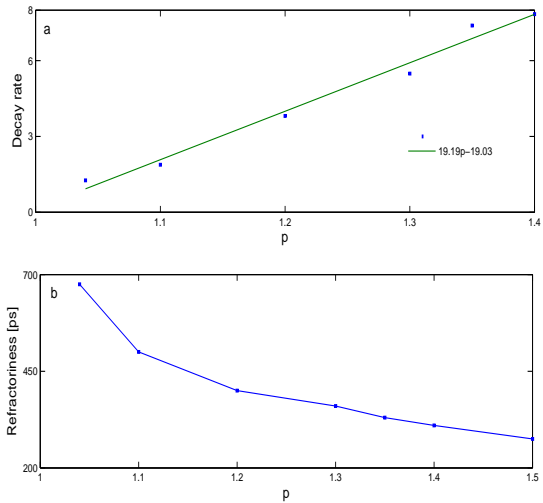}}}
\par}\caption{(a) The exponential decay rate as a function of $p$ obtained
for time intervals greater than the most probable time. The decay
rate is a linear function of $p$ for low $p$ values, as indicated by
the linear fit (solid line). (b) The most probable time interval,
the refractory time, as a function of $p$.   }
\end{figure}

For the case of a fully transparent mirror, the two face-to-face
lasers synchronize achronally, characterized by two dominant peaks
in the cross correlation function at $\pm\tau$. We find that the
statistics of the intervals between spikes, the refractory period
and the repetitive spiking pattern with optical delay time, $\tau$,
are not altered. However, when When the two face-to-face lasers are
configured to have different $p$ values, the lasers no longer
synchronize, and from a sample of the signals transmitted between
the lasers it is not obvious that one could not determine that the
two lasers are operating with two different p's and the values of
the two p's.  the The statistics of the spike intervals, however, is
altered.reveals tis information. Figure 4 5 depicts the distribution
of times between spikes when $p_1=1.1$ and $p_2=1.4$ where as
previously, the feedback strength for each laser is a few percent of
it's output power. Figure 4 5 indicates that the distribution for
each of the two lasers is a combination of the distributions of each
chaotic solitary laser, as shown in Figure 3a. The distribution
consists of two maxima related to the refractoriness connected to
$p_1$ and $p_2$. For time intervals greater than both refractory
times the distribution follows a Poisson distribution.  Similar
results are obtained for all combinations of p values lying in the
range of 1 to 1.5.  Thus a statistical analysis reveals information
about and differentiates between two sources operating with
different parameters.

\begin{figure}
\vspace{-0.5 cm} {\centering
\resizebox*{0.4\textwidth}{0.22\textheight}
{{\includegraphics[angle=0]{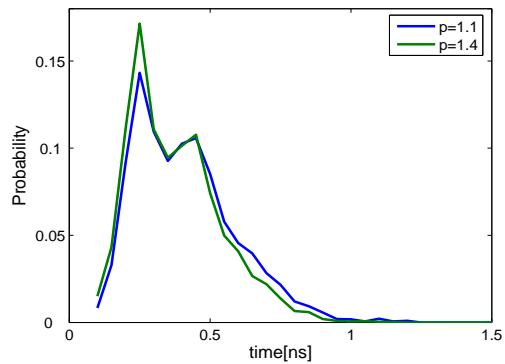}}}
\par}\caption{Two lasers in a face-to-face configuration corresponding to a fully
transparent mirror in the schematic of Figure 1. The first laser
operates with $p=1.1$ and the second laser with $p=1.4$ and the
coupling strength is a few $\%$ of the laser intensities. The
probability for the time intervals between spikes is presented for
each one of the lasers. The two maxima correspond closely to the
refractoriness of the solitary lasers operating with $p=1.1$ and
$1.4$ (see Figure 3).}
\end{figure}

Figure 4 5 indicates that although the two mutually interacting
lasers with differing $p$ values are not synchronized, the
distribution of the intervals between spikes of each individual
laser contains information about the parameter $p$ of the other
laser. More precisely, from the measurement of the distribution of
spiking time intervals containing the two refractory periods only
and using its own parameter $p$, each laser can deduce the parameter
$p$ of the other laser. This mechanism may play an important role in
neurobiology, where the statistical measure of the short intervals
among spikes may reveal information about the individual state of
the interacting neurons.

For the third scenario of a partially transmitting mirror, each
laser receives both self-coupling and mutual coupling signals, and
the two lasers synchronize isochronally with zero time lag. The
correlation coefficient for the intensity traces averaged over 200
100 ns long time segments, exceeds $0.95$ for $p>1.2$ and is around
$0.9$ for small $p$, where low frequency fluctuations \cite{29}
appear which are taken as partand are included in the calculation of
the statistics. In the following we investigate whether the
deviation from perfect synchronization is caused by a mismatch
between the timing of the spikes, by the difference in their heights
or by the background noise in our measurements.

Figure 5a 6a shows the histogram for the mismatch between the timing
of the spikes of the two lasers in the isochronal phase. The
histogram indicates that the most probable time difference between
the spikes of the two coupled lasers is zero since the average
difference between the timing is less than $25$ ps, our sampling
rate. The width of the histogram, limited by our detection
bandwidth, is extremely narrow ($< 80$ ps), and is not resolved by
our detection system.

We also examine the relative difference between the maximum
intensities of two correlated spikes in the two lasers, and the
histogram consisting of over $60,000$ pulses is presented in Figure
5b6b. Shown is the difference between the maximum of each spike for
each laser divided by the average of the spike amplitudes
$\Delta_{amp}/Ave_{amp}$. This result indicates that the average
relative difference between the maximum heights of temporally
correlated spikes is $\sim10\%$. It is possible that the mismatch
between the maximum intensities of these spikes is much lower, since
the maximum intensity measured is very sensitive to the precise
sampling time of the oscilloscope relative to the shape of a spike.
From these observations we cannot definitively determine the source
of the nonperfect synchronization, but it appears likely that it is
not in the spike timing but rather in the spike amplitude.

\begin{figure}
\vspace{-0.5 cm} {\centering
\resizebox*{0.45\textwidth}{0.35\textheight}
{{\includegraphics[angle=0]{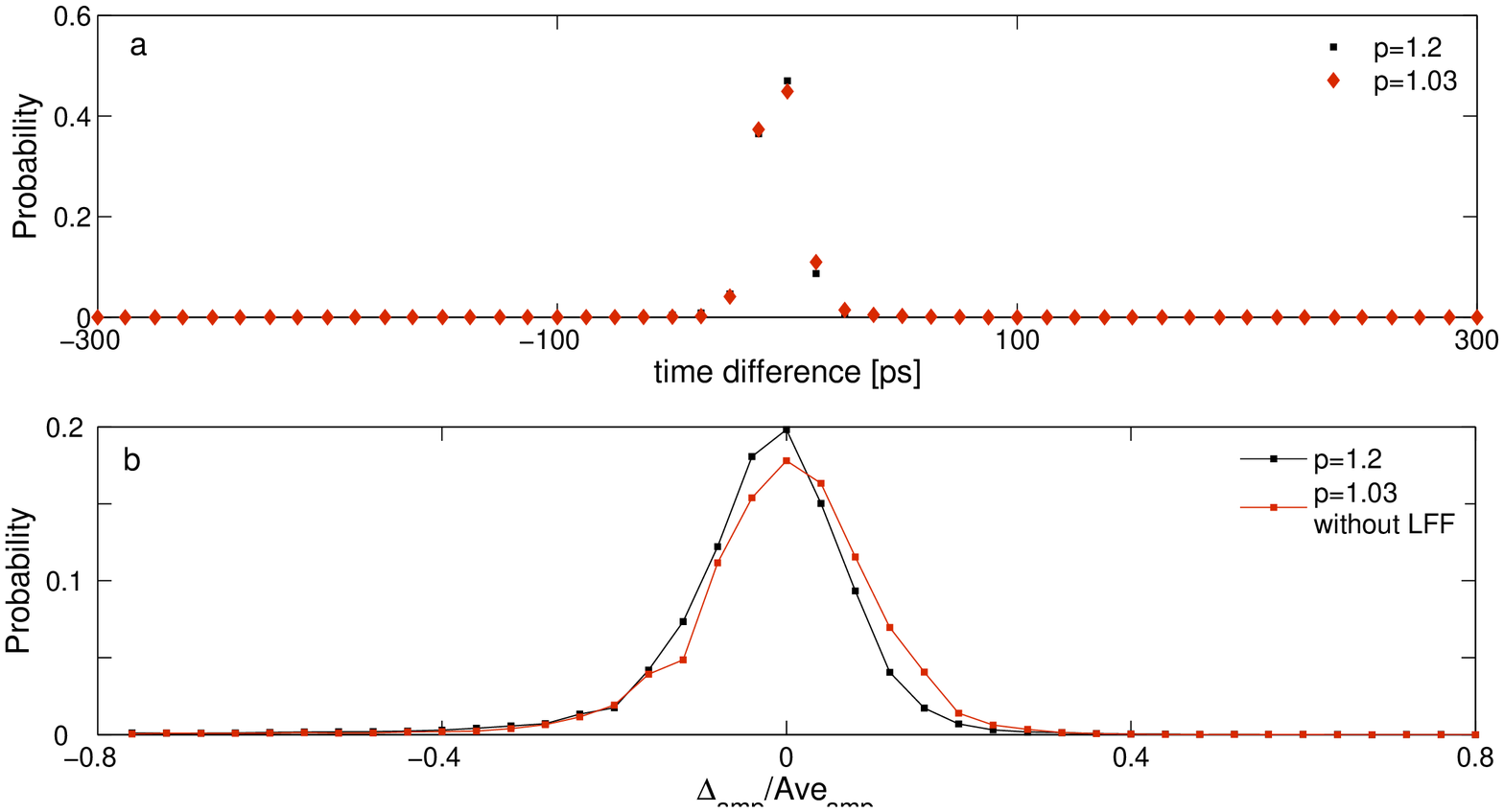}}}
\par}
\caption{Two lasers in a zero-lag isochronal phase with a partially
transmitting mirror in the schematic of Figure 1. (a) The histogram
of the mismatch in the timing of spikes of the two lasers, and (b)
the relative ratio between the maximum of the intensity of
temporally correlated spikes of the two lasers.  }
\end{figure}

Phenomenon similar to our observations have been found in the
communication of neurons, where immediately after the activation of
an action potential it is more difficult to excite a second spike.
Neural communication has been documented to have many features (such
as refractoriness, the repetitive form of the spiking pattern,
synchronization between spatially separated neuron groups, the spike
statistics \cite{21,22,23}) which are similar to those observed for
the chaotic lasers. Our demonstration that both the timing and
maximum intensities of spikes are extremely well synchronized with
zero time lag could have possible important implications for
corresponding neural system \cite{18,19}. One of the fundamental
problems in neuroscience, for example, is the question of how
information is encoded in the neuronal spike trains \cite{30}. Is
the information contained in an individual spike form or in the
interval between spikes, or is it the mean rate of spikes and timing
which matter \cite{31}? Traditionally it has been thought that most
of the relevant information was contained in the mean firing rate of
the neuron. It is clear, however, that an approach based on a
temporal average neglects all the information possibly contained in
the exact timing of the spikes and the statistical measure of the
short intervals among spikes may reveal information about the
individual state of the interacting neurons.. Recently more and more
experimental evidence has accumulated which suggests that a
straightforward firing rate concept based on temporal averaging is
too simplistic to describe neural information transfer. If, at each
processing step, neurons had to wait and perform a temporal average
in order to read the message, the reaction time would be
incompatibly long compared to experimental evidence.

In conclusion, our results for chaotic lasers show that individual
spiking laser units are able to generate irregular spike patterns
which become synchronized when two such units are coupled to each
other, without any time delay, although the transmission time can be
relatively long. Synchronization is maintained even on the time
scale of individual spike widths. For chaotic lasers, transmission
of information by the spiking pattern has been demonstrated, and the
repetitive bar-code pattern we observe has features which are useful
for communication applications of these signals.  The similarity of
the lasers to neural systems is noted, and it is possible that
complex neural systems can be effectively modeled by the much
simpler and much more flexible and well-controlled experimental
environment of coupled laser systems.


\end{document}